*Locality, Independence and the Pro-Liberty Bell*[1]

Huw Price[2]

Bell's Theorem requires the assumption that the values of hidden variables are statistically independent of future measurement settings. This *Independence Assumption* (IA) is normally taken for granted, though Bell himself sometimes considered relaxing it, as a way of defending locality. On balance he regarded such a move as even less attractive than non-locality, however—this despite the fact that he himself believed that non-locality conflicts with the orthodox interpretation of Special Relativity—for he thought that if we gave up IA we would have to abandon the belief that we are free.

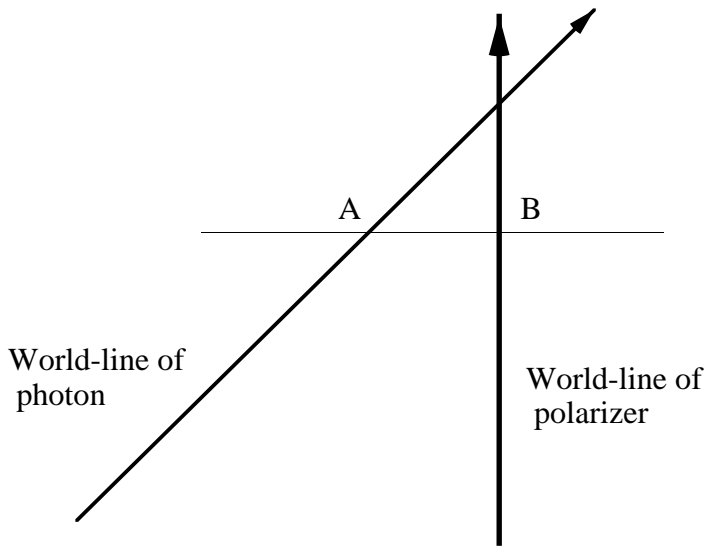

**Figure 1**

Few other commentators pay any explicit attention to the possibility of giving up IA. I'm not sure whether this is because Bell's concerns about freedom are widely shared, or simply because IA seems too obvious to challenge. At any rate, it is this unconventional path that I want to talk about here. The main point of the paper is to distinguish two quite different ways of giving up IA, which tend to be confused. I confused them myself for a long time, and as I'll explain, I think that others have, too, including Bell himself. But the

---

two proposals are really quite different. One of them, I think, is rightly dismissed; which means that unless we notice that there is an alternative, it is easy to make the mistake of thinking that IA cannot seriously be challenged. Approached from the right metaphysical perspective, however, the alternative looks like not only a serious possibility, but a very plausible hypothesis. In this case, however, it is easy to overlook the fact that this hypothesis is not what others who have considered rejecting IA have had in mind. I think that discussion of IA has been conducted at cross purposes, and my aim in this paper is to sort things out.

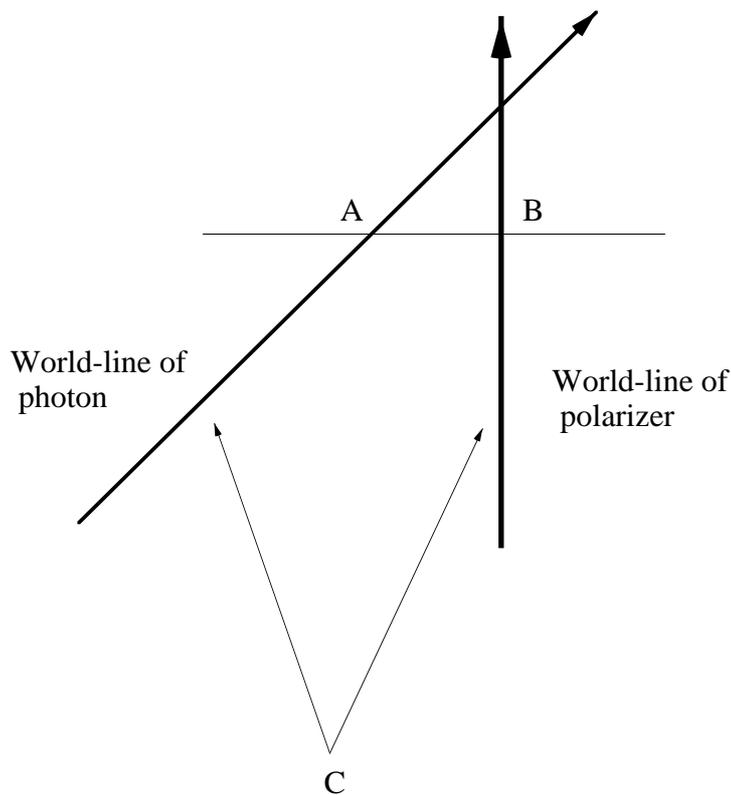

**Figure 2**

What are these two ways of challenging IA? Let's think first about IA itself, in terms of a simple example. Consider a photon approaching a polarising screen: IA is the assumption that the state of the photon is independent of the orientation of the polariser. If we assume that the setting of the polarizer is already fixed as the photon approaches, we can represent this assumption in terms of a simple world-line diagram, as in Figure 1. IA amounts to the assumption that the state of the photon at A is independent of the state of the polarizer at B.

What would it be for IA to fail? It would be for there to exist a correlation, or dependence, between A and B. If we assume, as special relativity and ordinary intuition both seem to recommend, that there is no direct causal connection between A and B—that neither is a cause of the other—then Reichenbach's (1956) *Principle of the Common Cause* tells us that we should expect the correlation between the two to be explained by joint correlations with an earlier common cause. In other words, we would expect that some factor C in the common past of the photon and the polarizer is responsible for their correlation. This, then, is the first way to relax IA—let's call it the *Common Past Hypothesis* (CPH).

CPH seems to have been mainly what Bell in mind when he considered relaxing IA. The following passage is the most explicit comment on the matter I know of in Bell's published work.

> It may be that it is not permissible to regard the experimental settings *a* and *b* in the analyzers as independent variables, as we did. We supposed them in particular to be independent of the supplementary [hidden variables] λ, in that *a* and *b* could be changed without changing the probability distribution ρ(λ). Now even if we have arranged that *a* and *b* are generated by apparently random radioactive devices, ... or by the Swiss national lottery machines, or by elaborate computer programmes, or by apparently free willed physicists, or by some combination of all of these, we cannot be sure that *a* and *b* are not significantly influenced by the same factors λ that influence [the measurement results] A and B. But this way of arranging quantum mechanical correlations would be even more mind boggling than one in which causal chains go faster than light. Apparently separate parts of the world would be conspiratorially entangled, and our apparent free will would be entangled with them. (Bell 1987, p. 154)

As this passage hints, Bell's main objection to CPH seems to have that the degree of determinism required appeared to him to be incompatible with free will. In a published interview he reinforces this impression, saying that in the analysis leading to Bell's Theorem

> it is assumed that free will is genuine, and as a result of that one finds that the intervention of the experimenter at one point has to have consequences at a remote point, in a way that influences restricted by the finite velocity of light would not permit. If the experimenter is not free to make this intervention, if that also is determined in advance, the difficulty disappears. (Davies and Brown 1986, p. 47)

The free will objection is a little puzzling, however. After all, a long tradition of philosophical compatibilism maintains that free will is compatible with classical Laplacian determinism; and yet surely the determinism involved here would be no more severe than that? (How could it be?) The compatibilists might be wrong, of course, but why prejudge the issue? More importantly, why take the issue to be relevant here, and not (say) as an objection to classical mechanics?

I suspect that Bell was influenced by two considerations. The first is a fatalist argument quite distinct from causal determinism: roughly, it is the thought that if the state of the incoming quantum object "already" reflects the measurement setting, then we are not free to choose that setting. The second factor is the thought that the required common cause would have to be something entirely new, something of a kind not presently envisaged in our view of the physical world. The conclusion that our actions are influenced by physical states of affairs of a previously unimagined kind could easily lead one to fatalism!

Fatalism aside, however, the latter factor supports another argument against CPH, namely that the hypothesis requires a universal mechanism of extraordinary scope and discrimination to maintain the correlations. Think of all the different ways in which measurement settings might be chosen. The mechanism would need to steer all these different physical pathways towards the same endpoint. No doubt these considerations could be made more precise, but the outline of the objection is clear: CPH needs to postulate a vast and implausible hidden substructure underlying what we presently think of as physical reality, in order to supply the required common causes.

If we reject CPH, is there any other way in which IA might fail? I think there is. In terms of our idealised example, it is the idea that there might simply be a correlation between A and B in virtue of the future interaction between the photon and the polarizer. In effect, it is the suggestion that we don't need to import the joint event C in order to explain the correlation between A and B, because the two systems involved already possess a joint event of their own: their joint future interaction.

Let us call this the *Common Future Hypothesis* (CFH). Obviously it escapes the most glaring objection to CPH, for it doesn't require anything additional in the world, over and above what we already take to be there. (Except perhaps the new hidden variables of the existing systems.) What's wrong with it? Why would anyone contemplating giving up IA prefer CPH to CFH?

As far as I can see, the only answer to this question rests on our commitment to what is effectively the Common Cause Principle: i.e., to the principle that remote correlations need an explanation in their joint past. But what is the status of this principle? Do we have any good reason to allow it to guide our thinking in this case?

My own view is that we do not. Briefly, I think a strong case can be made for the view that the Common Cause Principle is essentially macroscopic: roughly, it is a feature of the average behaviour of large numbers of physical systems, closely related to the thermodynamic asymmetry; and should provide no constraint on individual microscopic systems. From the individual photon's point of view, a correlation with a polarizer in virtue of a future interaction is no more problematic than a correlation in virtue of a past interaction.

A lot more needs to said about this. The temporal asymmetry of dependency is a very tricky issue, which I have discussed in other work. (See particularly Price 1992 and 1996, chs. 5–7.) But the upshot remains the same, I think: there is no particular justification for invoking the Common Cause Principle in this context; on the contrary, what we know about the nature of temporal asymmetry in general should lead us to favour a more symmetric view in microphysics.

Thus if one approaches the issue as I have, from the viewpoint of an interest in time asymmetry, CFH looks like the natural way to consider giving up IA. CPH simply perpetuates the very asymmetric assumption the challenge to which provides the motivation for questioning IA. Given this motivation for questioning IA, in other words, CPH doesn't even to seem to be an option, let alone the preferable option.

What I failed to appreciate for a long time is that not everyone who considers rejecting IA does so from this standpoint. Others approach the issue "directly," as I did at the beginning of this talk. Without a reason to challenge the Principle of the Common Cause, they take it for granted, and are thus led, quite properly, to CPH.

Bell himself was aware of both approaches to the issue, but it is not clear that he saw that they are distinct. He certainly knew of challenges grounded in general considerations about time asymmetry—for example, the work of Costa de Beauregard, which argues that the Bell correlations require what Costa de Beauregard calls zig-zag causality. (Bell referred to these ideas, I think, in talks he gave in Canberra in the early 1980s.) But when he mentions the idea of abandoning IA in print, it seems to be CPH that he has in mind. I have a letter from Bell in 1988, in which he first of all makes it clear that he is aware of what I have here been calling the CFH: he says, "I have not myself been able to make any sense of the notion of backward causation. When I try to think of it I lapse quickly into fatalism." (Bell 1988) But for his thoughts on the subject he then goes on to refer to his published discussion with Clauser, Horne and Shimony (in Bell, Clauser, Horne and Shimony 1985), where what is considered is CPH rather than CFH. (There is no talk of backward causation in that discussion.)

I suspect that Bell's failure to distinguish these two alternative approaches can be traced in part to his concerns about fatalism. Bell thought that they both conflicted with our intuitive assumption that experimenters are free to choose measurement settings, and that these settings are free variables. At one point in the discussion just mentioned, he characterises this assumption as the idea that 'the values of such variables have implications only in their future light cones. They are in no sense a record of, and do not give information about, what has gone before.' What is true is that both CPH and CFH deny this: both take the setting of the polarizer to be correlated with the earlier state of the photon. But this is where the similarity ends. Because the two views tell very different stories about what sustains this correlation, they may have very different implications concerning human freedom. In the case of CFH but not of CPH, it is plausible to argue that the relevant earlier states are under the control of the experimenter who chooses the measurement setting; and hence that the violation of this condition does not imply that the measurement setting are not free variables, in the most useful sense of that term.

As I noted earlier, there seem to be two quite separate arguments for fatalism: one, which applies only to CPH, turns on the idea that the hidden determinants of choice of measurement settings would be incompatible with free will; the other, which applies at least as much to CFH, at least at first sight, turns on the thought that if the photon's state is 'already' correlated with the future setting of the polarizer, then we can't really have any choice in the matter. I suspect that Bell didn't appreciate that these are quite different points.

My main concern in this paper has just been to draw some overdue attention to the difference between these two ways of relaxing IA. We should not confuse the proposal to abandon IA with CPH. There is a different way to abandon IA, which doesn't call for the massive and implausible supporting structure that CPH requires. True, the alternative is counterintuitive in a different way, but at this point it is arguable that the intuition concerned is simply mistaken—that it conflicts with an independently appealing position concerning the nature of temporal symmetry in the real world.

It may be helpful to finish by representing the issue in terms of a simplified map of logical space. Suppose we partition the set of possible worlds according to whether the Common Cause Principle holds at a microscopic level. In relation to this partition, where does the set of possible worlds in which IA fails fit in? Something like in Figure 3. (The top half of the large rectangle represents the possible worlds in which the Common Cause Principle holds, and the bottom half represents the worlds in which it fails. The shaded region represents the worlds in which IA fails.) This distribution accounts for the fact that discussion of the possibility of giving up IA has tended to proceed at cross purposes. Those who come to the discussion without challenging the Common Cause Principle confine their attention to the top half of the diagram, where the failure of IA is an insignificant and implausible possibility. Those who come to the discussion as I did, however—motivated by the thought that the Common Cause Principle is implausible at the microscopic level on T-symmetry grounds—confine their attention to the possible worlds in the bottom half, and see the rejection of IA as a natural and plausible move. From this point of view it seems that IA displays a deeply ingrained classical bias: an asymmetric intuition, for which there turns out to be no good basis. The way to correct this bias is not to entertain CPH—the view in the tip of the iceberg in Figure 3—which simply perpetuates the classical myth, but to tackle the problem directly; to acknowledge that physical systems may exhibit the same kind of brute correlation with their futures as they do with their pasts.

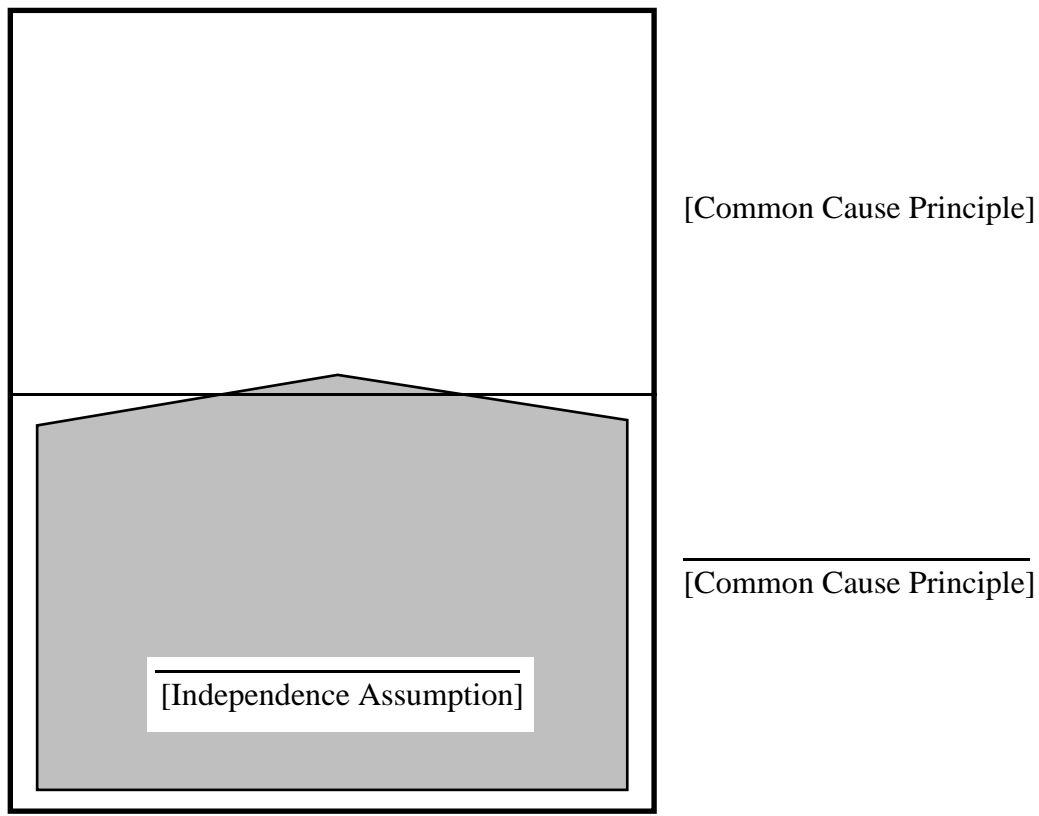

**Figure 3**

I have suggested that unlike CPH, CFH allows us to continue to think of experimental measurement settings as free variables in the old way—it simply makes certain earlier hidden states dependent variables. So it greatly reduces the tendency to think that rejecting IA requires a radically different view of the processes by which measurement settings are determined. What CFH does change is our view of measurement itself. It holds that measurements (and other phyical interactions, of course)

"create" or influence pre-existing states in quite a new way. But it is an orthodoxy that QM shows that measurement disturbs the observed system in some non-classical fashion. Different interpretations offer different accounts of what this non-classical process involves. The appropriate question is how the CFH account compares with the others on offer. The great advantage of CFH is that in rejecting IA, it offers a strategy for avoiding non-locality—and unlike CPH, it delivers at the same time another metaphysical bonus: it respects a temporal symmetry the need for which other views simply overlook. (I discuss the advantages of CFH at greater length in Price 1994, and 1996, ch. 9.)

I conclude that the issue of the status of Bell's Independence Assumption needs to be re-opened, with explicit attention to the role and status of the Common Cause Principle. Most previous discussions have missed the interesting point; or, in iceberg terminology, noticed only the point, and failed to see the potential of what lies below the waterline.